\documentclass[12pt,preprint]{aastex}
\begin{document}

\def\pdot {\dot P}
\def\Omdot {\dot \Omega}
\def\ltsima{$\; \buildrel < \over \sim \;$}
\def\lsim{\lower.5ex\hbox{\ltsima}}
\def\gtsima{$\; \buildrel > \over \sim \;$}
\def\gsim{\lower.5ex\hbox{\gtsima}}
\def\msole{~M_{\odot}}
\def\mdot {\dot M}
\def\cha {\textit{Chandra~}}
\def\cas {Cas~A~}
\def\xmm  {\textit{XMM-Newton~}}

\title{The X--ray Source  at the Center of \\ the  Cassiopeia A Supernova Remnant}
\author{S. Mereghetti}
\affil{Istituto di Fisica Cosmica G.Occhialini - CNR \\
v.Bassini 15, I-20133 Milano, Italy \\
sandro@ifctr.mi.cnr.it}
\author{A. Tiengo}
\affil{XMM-Newton Science Operation Center   \\
VILSPA ESA, Apartado 50727, 28080 Madrid, Spain \\
tiengo@xmm.vilspa.esa.es}
\author{G.L. Israel}
\affil{Osservatorio Astronomico di Roma    \\
v.Frascati 33, I-00040 Monteporzio  Catone, Italy \\
gianluca@mporzio.astro.it}

\begin{abstract}

We present the  first results of an \xmm  observation of the
central X--ray source in the \cas supernova remnant. The spectrum
can be fit equally well with an absorbed  steep power law
($\alpha_{ph}\sim$3,
N$_H\sim$1.5$\times$10$^{22}$ cm$^{-2}$) or
with a bremsstrahlung with temperature kT$\sim$2.4 keV and
N$_H\sim$10$^{22}$ cm$^{-2}$.
A blackbody model (kT$_{BB}\sim$0.7 keV) gives a slightly worse fit
and requires a  column
density N$_H\sim$5$\times$10$^{21}$ cm$^{-2}$ and an emitting area with
radius $\sim$0.3 km (for d=3.4 kpc).

A search for pulsations for periods longer than 150 ms gave negative
results. The 3$\sigma$ upper limits on the pulsed fraction
are $\sim$13\% for P$>$0.3 s and $\sim$7\% for P$>$3 s.

The overall properties of the central X--ray source in   \cas are
difficult to explain in terms of  a  rapidly spinning neutron
star with a canonical magnetic field of $\sim$10$^{12}$ G, and are
more similar to those of slowly rotating neutron stars such as
the Anomalous X--ray Pulsars.

\end{abstract}


\section{Introduction}

The long-sought   compact remnant born in    the \cas
supernova  explosion was discovered in the first light image
obtained with the \cha X--ray satellite (Tananbaum 1999).
For the following reasons there is little doubt that CXO J232327.8+584842 (CXO for brevity)
is indeed associated to \cas, the youngest ($\sim$320 yrs) supernova remnant observed
in the Galaxy (Ashworth 1980, Fesen et al. 1987).

First, the  X--ray flux of $\sim$10$^{-12}$ erg cm$^{-2}$ s$^{-1}$
(Pavlov et al. 2000, Chakrabarty et al. 2001)
and the strong upper limits on any optical/IR counterpart in the
small (1$''$ radius) \cha error circle (R$>$26.3, K$>$21.2, Ryan et al. (2001),
Kaplan et al. (2001)), imply for CXO an X--ray to optical
flux ratio greater than $\sim$800.
This immediately excludes any kind of non-compact galactic source as well
as accreting binary systems, unless the companion star is a very low mass object.
On the other hand the X--ray to optical flux ratio of CXO is similar to that
of radio pulsars and other neutron star candidates (see, e.g., Shearer \& Golden 2001,
Treves et al. 2000).

Second, CXO is located at an angular distance $\lesssim$7$''$ from the
site  of the \cas supernova explosion, that has been
recently re-determined
with good accuracy by a backward projection of the   expanding optical filaments
(Thorstensen et al. 2001).
For an age of  320 yrs and a distance of  3.4 kpc (Reed et al. 1995),
the implied transverse speed of the neutron star would be
$\sim$340 km s$^{-1}$, well within the range of the observed space velocities of radio pulsars
(Cordes \& Chernoff 1998).

Finally, the possibility of a background extra-galactic source has been dismissed
(Pavlov et al. 2000)
on the basis of  (a) the  very small probability of finding an AGN as bright as CXO
within  an area of only $\sim$0.05   arcmin$^2$, and (b) the   steep power
law X--ray spectrum of CXO.

Although the association with \cas is well established, the exact
nature of CXO remains puzzling. Deep radio observations
(McLaughlin et al. 2001) were unable to detect any  source, either
pulsed (upper limit of 1.3 mJy at 1.4 GHz) or unpulsed   (u.l. of
40 and 6 mJy at 1.3 and 4.4 GHz, respectively). A radio/X--ray
synchrotron nebula could indicate the presence of a rapidly
rotating, magnetized neutron star, even when beam orientation
effects prevent its direct observation as a radio pulsar.
However, no radio   or X--ray (Murray et al. 2001) nebula has
been found around CXO.
 
The X--ray spectrum of CXO is not well constrained.
Equally good fits to the \cha data were obtained   with simple power law,
blackbody and bremsstrahlung  models 
(Pavlov et al. 2000, Chakrabarty et al. 2001, Murray et al. 2001).
Power law fits give a steep photon index $\alpha_{ph}$$\sim$3-4 and
a column density  N$_H$$\sim$(1.5-2.3)$\times$10$^{22}$ cm$^{-2}$,
larger than the estimated  value  of the interstellar
absorption in the direction of \cas (Keohane et al. 1996).
Young neutron stars emitting in the X--ray band by non-thermal magnetospheric
processes have   harder   spectra ($\alpha_{ph}$$\sim$1.5)
and higher luminosity  (L$_X$$\sim$10$^{35}$-10$^{36}$ erg s$^{-1}$) than CXO.
The fits with a blackbody model give   kT$_{BB}$$\sim$0.5 keV  and
N$_H$$\sim$10$^{22}$ cm$^{-2}$, but they imply an emitting area
that, for any reasonable assumed distance, is far too small to be consistent
with the surface of a neutron star.
This discrepancy is not uncommon when fitting cooling neutron stars
with simple blackbody spectra.
Larger emitting areas, more consistent with the neutron star
dimensions,  are generally  obtained when more appropriate models for
the thermal emission from neutron star atmospheres are used (see, e.g., Zavlin et al. 1998, 1999).
However, Pavlov et al. (2000) showed that in the case of \cas also fits with neutron star atmosphere models
imply emission from only a small fraction of the   star surface.

A re-analysis of archival Einstein and ROSAT observations showed no evidence for
long term variability (Pavlov et al. 2000).
Search for X--ray pulsations have been carried out with \cha by Chakrabarty et al. (2001)
with negative results, and later by
Murray et al. (2001)  who reported a possible periodicity at 12 ms,
but with a very low statistical significance.
If this period were confirmed,
it would be  difficult to explain the lack of a strong pulsar wind nebula around CXO, unless
its magnetic field is rather low ($\lesssim$5$\times$10$^{10}$ G).

Here we report the results of an observation of CXO
performed with the \xmm satellite on July 27$^{th}$, 2000 as part of the
Performance Verification program.

\section{Observations}

The data reported here were obtained with the
European Photon Imaging Camera (EPIC) instrument on the \xmm satellite.
EPIC consists of two MOS CCD detectors (Turner et al. 2001) and a PN
CCD instrument (Str\"{u}der et al. 2001), giving a total collecting
area $\gtrsim$ 2500 cm$^2$ at 1.5 keV. The mirror system offers an
on-axis FWHM angular resolution of 4-5$''$ and a field of view of 30$'$ diameter.

The \xmm observation of \cas started  at 6:45 UT of July
27$^{th}$, 2000   and lasted about 11 hours. The telemetry
saturation due to the high overall counting rate from the
supernova remnant caused a non-continuous  operation of the CCDs,
yielding effective exposures of 9219 s in the  PN,  and
23573 s and 23594 s in  the MOS1 and MOS2 detectors, respectively.
The satellite was
pointed at the center of the remnant (RA: 23$^h$ 23$^m$ 25$^s$,
Dec: 58$^{\circ}$ 48$'$ 20$''$ (J2000)), resulting in the
detection of CXO very close to the center of the EPIC field of
view (off-axis angle of 0.52$'$). During this
observation the thin filter was used. The MOS cameras were
operated in Large Window mode to limit the source photon pile-up.
The background corrected count rates of CXO were  0.25$\pm$0.02
and  0.13$\pm$0.01 counts s$^{-1}$, respectively in the PN and in each MOS.

The data were analyzed with version 5.2 of the \xmm Science Analysis System.
The counts for the spectral analysis were extracted in the three EPIC cameras
from circles with 10$''$ radii centered at the \cha position of CXO.
The corresponding background corrected count rates  were  0.25$\pm$0.02
and  0.13$\pm$0.01 counts s$^{-1}$, respectively in the PN and in each MOS.
Such a small radius, enclosing only about 50\% of the counts from a point source,
was chosen to avoid contamination from bright nearby filaments due to the
emission from the supernova remnant
(see the \cha images of Murray et al. (2001) and Chakrabarty et al. (2001)).
All the spectral fits    reported below were obtained with the appropriate
response matrices that take into account the dimensions of the
extraction region. The derived fluxes are  corrected for the fraction
of counts outside the 10$''$ circle.

CXO is relatively faint compared to the
non-uniform, diffuse emission due to \cas. We estimate that the background
accounts for $\sim$75-80\%
of the counts within the adopted extraction radius
(14050 for the PN and 13320 for the MOS1).
Therefore, we devoted  particular attention  to the choice of the region
for the extraction of the background spectra.
Based on the high resolution \cha images,
we tried different regions with a surface brightness similar to that at the position
of CXO and avoiding bright filaments.
We found that the resulting CXO spectra are somewhat
dependent on the chosen background that can lead to the presence of residuals
near the energies of the strongest   \cas emission lines.
Such residuals in the background
subtracted spectra were minimized using the background from a rectangular
($\sim$20$\times$15 arcsec$^2$)   region to the East of CXO,
roughly corresponding to the eastern half of that
used in the \cha analysis by Chakrabarty et al. (2001).
The spectra discussed below were obtained with this choice of the background
region.

For   the PN spectra we used photons   corresponding to single and double events,
after  checking that the individual     single and double
events spectra gave the same results within the uncertainties.
The energy channels were
grouped to have at least 30 net counts per bin, and the fits were performed in
the 0.6-10 keV energy range.
The results of power-law, blackbody and bremsstrahlung fits are
summarized in Table~1.

The same extraction and binning criteria were applied to the two
MOS cameras. While the results obtained with MOS1 agree with the PN ones
(see Table~1), the MOS2 fits gave different parameters and worse $\chi^2$ values.
We found that this discrepancy is due to the fact that at the time of
the \xmm observation of \cas   the offset calculation for the MOS cameras was
still based on an imperfect on-board algorithm  (only later in the mission
better fixed tables were implemented). This problem can cause spurious event
detections in channels close to the low energy
threshold and also produces an offset of the energy scale in particular regions
of the CCD.
By carefully examining the MOS2 image and comparing spectra of different regions,
we verified that in this observation such an effect occurred just at the
position of CXO in the MOS2. Since these effects cannot be  easily
corrected, we did not consider  the MOS2 spectra in our further analysis.

To better constrain the spectral parameters, we fitted jointly the PN and MOS1
data (Table~1).
The power law and bremsstrahlung models give better fits to the data, while
the fit with a blackbody (Fig.1) yields higher $\chi^2$ values.
For comparison with the results of Murray et al. (2001),
we present in Table~1 also the
values obtained by fixing N$_H$=1.1$\times$10$^{22}$ cm$^{-2}$,
and by adding a power law to the blackbody model.
For all the models, the observed 0.1-10 keV flux from the
joint PN-MOS1 fits is of the order of
$\sim$2$\times$10$^{-12}$ erg cm$^{-2}$ s$^{-1}$.
The systematic difference between the PN and MOS1 fluxes
must be ascribed to the uncertainties in the relative calibration of the
two detectors.
The errors on the luminosities reported in Table 1 have been
computed taking into account the uncertainties of the spectral
parameters and absorption (90\% c.l. for two interesting parameters).

For the timing analysis     we extracted all the counts with energy
in the 1 to 6 keV energy range from circles
with radius 10$''$ centered on the CCO position. This  yielded respectively
12276, 12652 and 12511 counts for the PN, MOS1 and MOS2 detectors.
The upper limits reported below take into account that the background,
estimated from nearby regions, accounts for $\sim$75\% of these
counts.
After correction of the arrival times
to the Solar System barycenter, a Fourier analysis of the background subtracted
light curves was performed using the method described in Israel \& Stella (1996).
The frequency resolution was 2.46$\times$10$^{-5}$ Hz.
The frame integration times of 0.073 and 0.9 s, respectively in the PN and MOS cameras,
set the upper limits to the explored frequency ranges.
No significant periodicity was found. The upper limits on the pulsed fraction,
computed according to  Vaughan et al. (1994), are shown in Fig.~2 as a function of the frequency.
The curves refer to the 3$\sigma$ upper limit on the source
pulsed fraction in the 1-6 keV energy range
assuming a sinusoidal modulation.
Above 0.4 Hz only the PN data could be used, while at lower frequencies we could
sum the power spectra from the three instruments, thus improving the
sensitivity as shown by the lower curve of Fig.~2.
Also separate searches in the soft (1-2 keV) and hard (2-6 keV) energy
ranges gave negative results. The corresponding upper limits are about a factor two higher than
those shown in Fig.~2.

\section{Discussion}

Our spectral results can be compared with those derived from the \cha observations.
In the  power law case, our best fit values for $\alpha_{ph}$ and N$_H$  are consistent with the
values found by Chakrabarty et al. (2001), but not with the more recent findings of
Murray et al. (2001) who obtain a softer and more absorbed spectrum.
The temperature values that we find for the bremsstrahlung and blackbody models are   higher
than the \cha values, and the corresponding column densities slightly smaller.
The blackbody normalization of our fits implies an emitting area with radius of 0.32d$_{3.4kpc}$ km,
smaller than that inferred with \cha.
To check whether the discrepancy with respect to the \cha results might be
explained by the higher background contribution in the EPIC spectral
extraction region, we also performed
an analysis with smaller source extraction regions, down to a radius of 5$''$ (always
using the appropriate response matrices).
The derived best fit parameters did not vary significantly.

The above mentioned difficulties in interpreting the X--ray emission from CXO
in terms of a young active pulsar
are therefore confirmed by our \xmm results.
An independent knowledge of the expected column density could help to
discriminate among the various spectral models.
Values in the range (0.9-1.3)$\times$10$^{22}$ cm$^{-2}$ have been derived by Keohane et al. (1996),
with significant variations across the remnant
and with relatively   smaller absorption at the position of CXO.
On this basis, the thermal bremsstrahlung fit is the one giving the most
consistent N$_H$ value.
The low absorption, as well as the higher $\chi^2$ value,
suggests that
the simple blackbody fit is not
adequate. The trend of the   residuals  shown in the
lower panel of Fig.~1 (see also Fig.3 of Murray et al. (2001))
probably indicates that atmosphere models more complex
than simple blackbody spectra are required.
 
The soft spectrum and high X--ray to optical flux ratio make
CXO similar to the Anomalous X--ray Pulsars (AXPs, Mereghetti 2001a).
These objects have periods in the 6-12 s range, steadily increasing on
timescales of $\sim$10$^3$-10$^5$ years, and two (possibly three) of them are located
at the center of shell-like supernova remnants. Similar properties (except for the periodicity)
are shared by the   X--ray sources at the center of the supernova remnants
Puppis A (Petre et al. 1996, Zavlin et al. 1999),
G290.5+10.0 (Mereghetti et al. 1996, Zavlin et al. 1998),
RCW 103 (Gotthelf et al. 1999), and G266.1-1.2 (Slane et al. 2001, Mereghetti 2001b,
Pavlov et al. 2001a).
The bolometric blackbody luminosity and temperature of all these sources are plotted in Fig.~3, where
for comparison we have also included two classes of older neutron stars:
three middle aged radio pulsars from which thermal emission from internal cooling has been observed
(PSR B1055-52, PSR B0656+14, and Geminga; see, e.g., Becker \& Tr\"{u}mper 1997) and several
nearby   isolated neutron stars
(Treves et al. 2000, and references therein).
For all the sources in which also a non-thermal spectral component
is present, we have plotted in Fig.~3 only the bolometric luminosity of the
blackbody component.

The AXPs have on average a higher luminosity than the central compact objects (CCOs)
in supernova remnants, but their blackbody temperatures are very similar.
Although the majority of AXPs and CCOs have quite stable luminosity,
there are two interesting variable sources that might be intermediate objects  linking
these two classes.
The candidate AXP AX J1845--03 has been observed since 1997
at a flux level corresponding to L$_X$$\lesssim$10$^{34}$ erg s$^{-1}$
(Torii et al. 1998, Vasisht et al. 2000),
at least ten times smaller than at the time of its discovery.
1E~1614--5055, the CCO in RCW~103, aready known to be variable between
F$_{0.5-2 keV}$$\sim$3 and $\sim$50$\times$10$^{-13}$ erg cm$^{-2}$ s$^{-1}$
(Gotthelf et al. 1999) and with a possible $\sim$6 hr periodicity
(Garmire et al. 2000a), has recently been
detected with \cha in an even higher    state
(Garmire et al. 2000b).

Alpar (2001) proposed a unified interpretation of   the classes of objects
shown in Fig.~3, in which they are all assumed to be neutron stars with ordinary
magnetic fields (10$^{11}$-10$^{13}$ G), thus avoiding  the need for super-critical
magnetic fields required by the "Magnetar" models for AXPs (e.g., Thompson \& Duncan 1996).
According to Alpar (2001),  isolated neutron stars experience a mass inflow due to
the formation of a residual accretion disk from fallback material after the supernova explosion.
This causes an evolutionary phase in the propeller regime, before the neutron star
turns on as a radio pulsar or as an AXP, depending on the initial conditions in terms
of rotational period, magnetic field and mass in the residual disk
(see also Chatterjee et al. 2000 for a similar model).
The dim, nearby neutron stars in the lower left corner of Fig.~3 are interpreted
as examples of sources in the propeller regime, in which the X--ray emission is
from cooling powered by internal friction. The CCOs would be the propeller objects
with the highest mass inflow, surrounded by an optically thick corona and might evolve
into AXP (Alpar 2001).

Searches for periodicities in the CCOs extending to low frequencies are therefore
relevant, since some of these objects might have periods similar to those of the AXP.
In this respect, we note that, while the CCO in G290.5+10.0 has a fast period
of 424 ms (Zavlin et al. 2000) and might be a ''radio-quiet'' pulsar
similar to Geminga, the fast periodicity  at 75 ms reported in the Puppis A CCO (Pavlov  et al.  1999)
has not been confirmed by more accurate \cha observations (Pavlov et al. 2001b).
For frequencies $\nu<$6 Hz, our limits on the pulsed fraction of CXO are  smaller
than those previously obtained  with \cha (Chakrabarty et al. 2001, Murray et al. 2001).
Only one of the AXPs has a pulsed fraction as small as 7\% (see, e.g., Israel et al. 2001),
our upper limit for long periods.

\section{Conclusions}

We have reported the first results of an \xmm observation of CXO,  the interesting
X--ray point source at the center of the \cas supernova remnant.
The large collecting area of the \xmm instruments has allowed us to derive   strong
upper limits of $\sim$13\% for P$>$0.3 s and $\sim$7\% for P$>$3 s  on the pulsed fraction (3$\sigma$).

The spectral analysis confirmed the very soft spectrum  of CXO , but yields,
in the case of thermal models,   higher temperatures than those measured
earlier with \cha. The source luminosity does not show evidence of long term variations
when compared with previous measurements.

As for other candidate neutron stars in supernova remnants
older than \cas,
the overall properties of CXO are difficult to explain in terms of  a  rapidly spinning neutron star
with a canonical magnetic field of $\sim$10$^{12}$ G,
and are more similar to those of slowly rotating neutron stars such as the AXPs.

\acknowledgments

We acknowledge the finantial support of the Italian Space Agency.
This work is based on observations obtained with \xmm,
an ESA science mission with instruments and contributions directly
funded by ESA Member States and NASA.

\begin{deluxetable}{lcccccc}
\tablecaption{Results of spectral fits}
\tablehead{
\colhead{Model}               &
\colhead{N$_H$$^a$} &
\colhead{$\alpha_{ph}$$^a$} &
\colhead{kT$^a$}    &
\colhead{Flux$^b$} &
\colhead{L$_{X}^c$} &
\colhead{$\chi^2_{red}/dof$}
}

\startdata
         & (10$^{22}$ cm$^{-2}$) &   & (keV) & (erg s$^{-1}$cm$^{-2}$) &
(10$^{33}$erg s$^{-1}$) & \nl
\hline
           PN       &        &                &             &   &
&     \nl
Power law         &   1.47$^{+0.34}_{-0.27}$ & 2.85$^{+0.31}_{-0.27}$
&    & 2.3 10$^{-12}$  &  51$^{+101}_{-25}$ & 1.09/30   \nl
Blackbody         &   0.46$^{+0.19}_{-0.15}$ &      &  0.75$\pm
0.06$              &  2.0  10$^{-12}$   &  3.5$\pm 0.3$ &   1.63/30
\nl
Bremsstr.         &   1.04$^{+0.23}_{-0.18}$ &      &
2.64$^{+0.65}_{-0.50}$ & 2.2  10$^{-12}$  &  8.0$^{+1.9}_{-1.3}$   &
1.13/30   \nl
                  &      &    &                &    &   &   \nl
\hline
    MOS1         &                        &   &
&                 &      &             \nl
Power law        & 1.53$^{+0.42}_{-0.33}$ & 3.16$^{+0.46}_{-0.39}$ &
&  1.9 10$^{-12}$  & 92$^{+364}_{-69}$ &  1.04/29   \nl
Blackbody        & 0.47$^{+0.18}_{-0.17}$ &   & 0.65$\pm 0.07$
&  1.67 10$^{-12}$ & 3.1$^{+0.5}_{-0.3}$ &  1.27/29   \nl
Bremsstr.        & 1.06$^{+0.27}_{-0.13}$ &   & 1.98$^{+0.60}_{-0.42}$
&  1.8 10$^{-12}$  & 8.2$^{+3.5}_{-2.0}$  &  1.03/29   \nl
                 &      &    &                &    &     & \nl
\hline
 PN+MOS1  &      &    &                &    & &      \nl
Power law   &   1.47$^{+0.25}_{-0.21}$  & 2.96$^{+0.25}_{-0.23}$  &  &
2.1 10$^{-12}$ & 60$^{+63}_{-29}$ & 1.07/61 \nl
Blackbody    &   0.46$^{+0.14}_{-0.12}$  &       &  0.70$\pm$0.05    &
1.8 10$^{-12}$ &  3.2$^{+0.3}_{-0.2}$ & 1.49/61  \nl
Bremsstr.     &  1.03$^{+0.17}_{-0.14}$  &  &  2.36$^{+0.44}_{-0.36}$ &
2.0 10$^{-12}$ &  7.9$^{+1.5}_{-1.2}$ & 1.11/61 \nl
Power law     &   1.1 (fixed)  & 2.60$\pm$0.11  &  & 2.1 10$^{-12}$ &
24.3$^{+5.5}_{-4.3}$ & 1.20/62     \nl
Blackbody    &    1.1 (fixed)   &   & 0.60$\pm$0.03 & 1.7 10$^{-12}$ &
4.2$^{+0.1}_{-0.1}$ & 2.00/62 \nl
PL+bbody & 1.1 (fixed) & 2.62 $^{+0.33}_{-0.32}$ & 0.62
$^{+0.18}_{-0.12}$ & 2.1 10$^{-12}$ & 19.5$^{+22.8}_{-9.9}$  & 1.09/60
\nl

\enddata

$^a$ Errors at 90\% confidence level for a single interesting parameter

$^b$ Absorbed flux 0.1-10 keV

$^c$ 0.1-10 keV luminosity (d=3.4 kpc) corrected for absorption.

\end{deluxetable}

\clearpage

\figcaption{EPIC PN and MOS1 spectra of CXO fitted with a blackbody model.
\label{fig1}}

\figcaption{Upper limits (3$\sigma$) on the CXO pulsed fraction (semiamplitude of a sinusoidal modulation
after background subtraction).
The curve for $\nu<$0.4 Hz
was obtained from the sum of MOS1, MOS2 and PN data, while at higher frequencies only the
PN data could be used.
\label{fig2}}

\figcaption{''Color-Magnitude'' diagram for different classes of neutron stars: Anomalous X--ray
Pulsars (stars), supernova remnant Central Compact Objects (squares),
nearby Isolated Neutron Stars (circles) and radio pulsars (diamonds). For the luminosity of AXP and   radio
pulsars only the thermal spectral component is considered. To allow the comparison between the
different objects we plotted the temperatures inferred from pure blackbody fits; the use of
atmosphere models would move all the points to lower temperatures by a factor $\sim$2-3.
The luminosity error bars reflect the distance uncertainties. The two positions
for RCW~103 and AX J1845--03 indicate the observed variability range.
\label{fig3}}


\clearpage
\epsscale{.7}
\plotone{f1.ps}

\clearpage
\epsscale{0.9}
\plotone{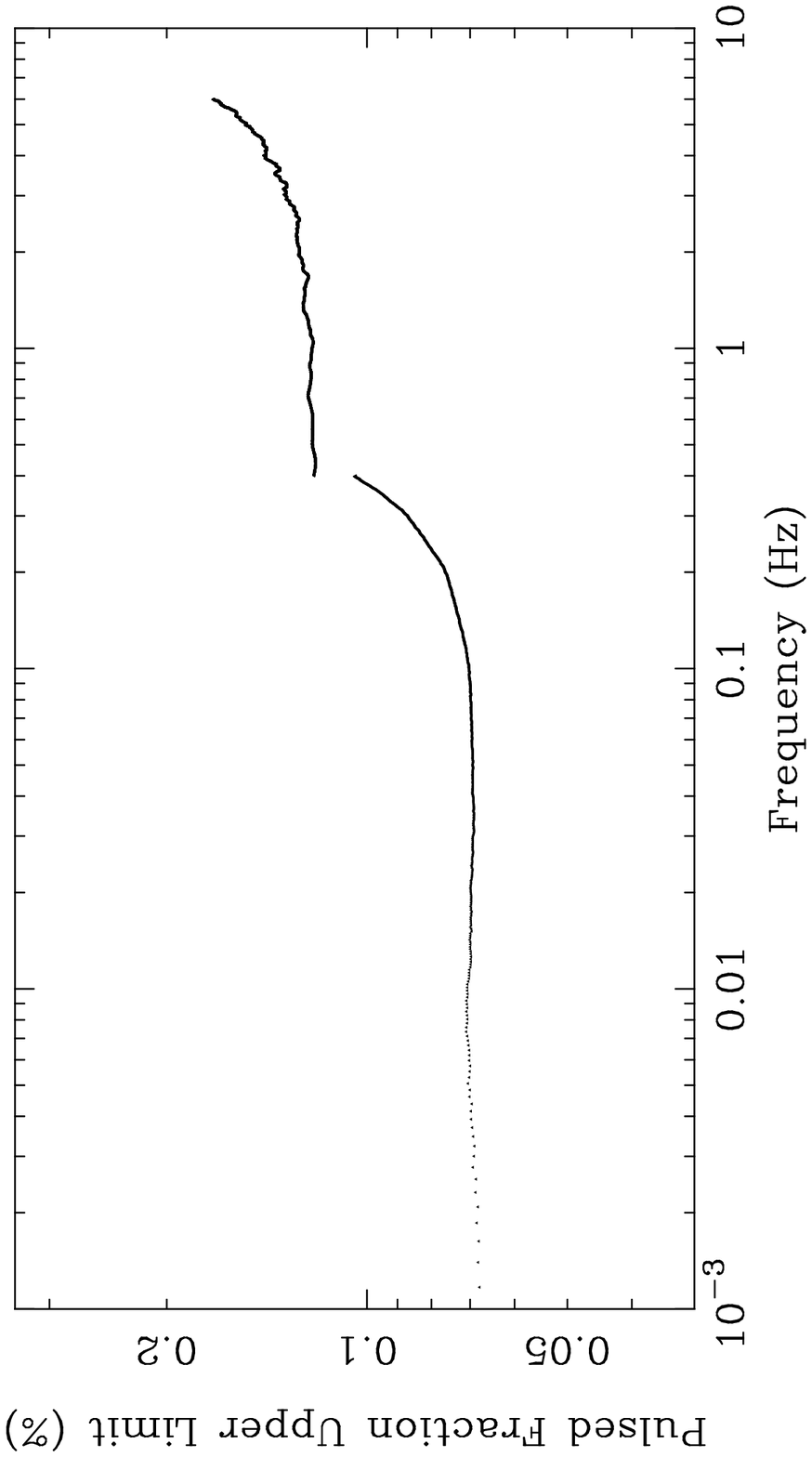}

\clearpage
\epsscale{.8}
\plotone{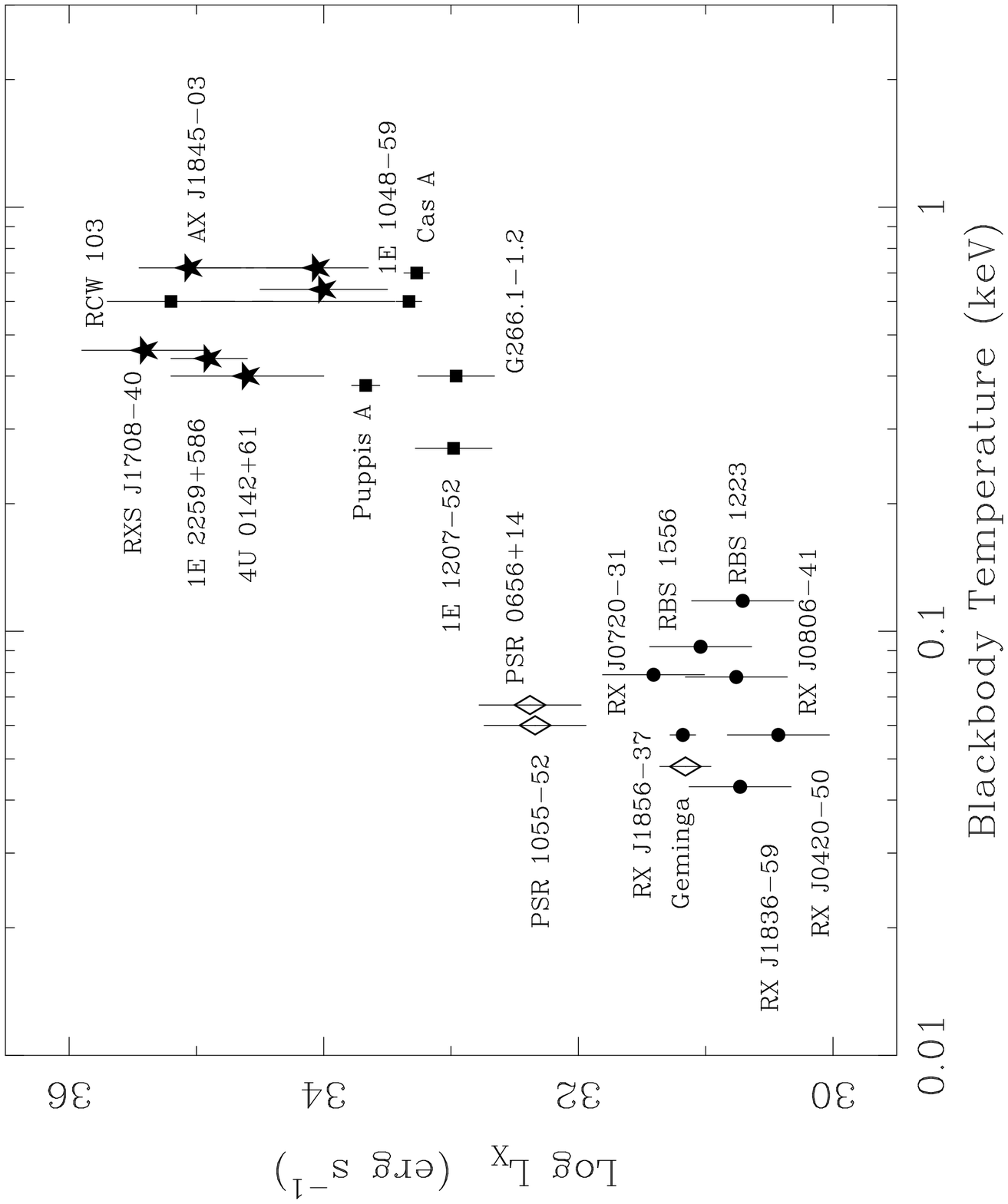}


\begin{references}
\reference{} Alpar, A. 2001, \apj, 554, 1245
\reference{} Ashworth W.B. 1980, J. Hist. Astron., 11, 1
\reference{} Becker W. \& Tr\"{u}mper J. 1997, A\&A, 326, 682
\reference{} Chakrabarty D. et al. 2001, \apj, 548, 800
\reference{} Chatterjee P., Hernquist L. \& Narayan R. 2000, \apj, 534, 373
\reference{} Cordes J.M. \& Chernoff D.F. 1998, \apj, 505, 315
\reference{} Fesen R.A., Becker R.H. \& Blair W.P. 1987, \apj, 313, 378
\reference{} Garmire G., et al. 2000a, IAU Circ. n.7350
\reference{} Garmire G., et al. 2000b, AAS HEAD Meeting 32, 32.11
\reference{} Gotthelf E.V., Petre R. \& Vasisht G. 1999, \apj, 514, L107
\reference{} Israel G.L. \& Stella L. 1996, \apj, 468, 369
\reference{} Israel G.L., Mereghetti S. \& Stella L. 2001, in Proc.   Miniworkshop on Soft Gamma Repeaters
and Anomalous X-ray Pulsars, Rome, October 2000,  eds. M.Feroci \& S.Mereghetti, Mem. S.A.It., 
in press, astro-ph/0111093
\reference{} Kaplan D.L., Kulkarni S.R. \& Murray S.S. 2001, \apj, 558, 270
\reference{} Keohane J.W., Rudnick L. \& Anderson M.C. 1996, \apj, 466, 309
\reference{} McLaughlin M.A. et al. 2001, \apj, 547, L41
\reference{} Mereghetti S., Bignami G.F.  \& Caraveo P.  1996, \apj, 464, 842
\reference{} Mereghetti S. 2001a, in Proc.  ''The Neutron Star - Black Hole Connection'',
eds. C.Kouveliotou, J. van Paradijs \& J.Ventura,  NATO ASI Ser., Dordrecht Kluwer,
in press, astro-ph/9911252
\reference{} Mereghetti S. 2001b, \apj, 548, L213
\reference{} Murray S.S. et al. 2001, \apj, in press
\reference{} Pavlov G.G., Zaviln V.E.  \& Tr\"{u}mper J.  1999, \apj, 511, L45
\reference{} Pavlov G.G., et al. 2000, \apj, 531, L53
\reference{} Pavlov G.G., et al. 2001a, \apj, 559, L131
\reference{} Pavlov G.G., Sanwal D., Garmire G.P. \& Zavlin V.E. 2001b,  
in ''Neutron Stars in Supernova Remnants'', ASP Conf. Proceedings, P.O.Slane \& B.M.Gaensler eds., 
in press, astro-ph/0112322
\reference{} Petre R., Becker C.M. \& Winkler P.F. 1996, \apj, 465, L43
\reference{}  Reed J.E. et al. 1995, \apj, 440, 706
\reference{} Ryan E., Wagner R.M. \& Starrfield S.G. 2001, \apj, 548, 811
\reference{} Slane P. et al. 2001, \apj, 548, 814
\reference{} Shearer A. \& Golden A. 2001, \apj, 547, 967
\reference{} Str\"{u}der L. et al. 2001, A\&A 365, L18
\reference{} Tananbaum H. 1999, Iau Circ. 7246
\reference{} Thompson C. \& Duncan R.C. 1996, \apj, 473, 322
\reference{} Thorstensen J.R., Fesen R.A. \& van den Bergh S. 2001, \aj, 122, 297
\reference{} Torii K. et al. 1998, \apj, 503, 843
\reference{} Treves A., Turolla R., Zane S. \& Colpi M. 2000, PASP 112, 297
\reference{} Turner M.J.L. et al. 2001, A\&A 365, L27
\reference{} Vasisht G., Gotthelf E.V., Torii K. \& Gaenlser B.M. 2000, \apj, 542, L49
\reference{} Vaughan B.~A., van der Klis M., Wood K.~S., et al., 1994, \apj,  435, 362
\reference{} Zaviln V.E., Pavlov G.G. \& Tr\"{u}mper J. 1998, A\&A, 331, 821
\reference{} Zaviln V.E., Tr\"{u}mper J. \& Pavlov G.G. 1999, \apj, 525, 959
\reference{} Zaviln V.E., Pavlov G.G., Sanwal D. \& Tr\"{u}mper J. 2000, \apj, 540, L25
\end{references}
\end{document}